\documentclass[10pt]{article}
\usepackage{graphicx,color}
\usepackage{url}

\title{The ratio  of $e^{\pm}p$ scattering cross  sections
predicted from the global fit of elastic $ep$ data}

\author{W. M. Alberico$^1$, S. M. Bilenky$^{2,3}$, C. Giunti$^1$,  K. M. Graczyk$^{1,4}$\footnote{graczyk@to.infn.it} \\ \\
$^1$\textit{\small Dipartimento di Fisica Teorica, Universit\`a di Torino and INFN}, \\
\textit{ \small Sezione di Torino, I--10125 Torino, Italy} \\
$^2$\textit{ \small BLTP, JINR, RU-141980 Dubna, Russia }\\
$^3$\textit{\small Physics-Department E15, Technische Universit\"{a}t M\"{u}nchen, D-85748 Garching, Germany}\\
$^4$\textit{\small Institute of Theoretical Physics, University of Wroc\l aw, pl. M. Borna 9, 50-204, Wroc\l aw, Poland}}

\begin{document}

\maketitle

\begin{abstract}
We present predictions for the value of the cross section ratio
$\sigma(e^+p \to e^+p )/\sigma(e^-p \to e^-p )$, determined from our
fit of the elastic $ep$ cross section and polarization data. In this fit we took into account the
phenomenological two-photon exchange dispersive correction.
 The cross section ratios which are expected to be measured by the
VEPP-3 experiment are computed. The kinematical region which will be covered by the E04-116
JLab experiment is also considered. It is shown that for both experiments the
predicted cross section ratios deviate from unity  within more than $3\sigma$.
\end{abstract}



\section{Introduction}

The electric and magnetic proton form factors can be obtained from the measurement of the elastic
$ep\to ep$ cross section. The data are usually analyzed via the Rosenbluth separation technique.
This amounts to write the reduced cross section, in the one-photon-exchange approximation, in units of the Mott cross section, as follows:
\begin{equation}
\label{sigma_reduced}
\sigma_{R,1\gamma}(Q^2,\epsilon) = G_{Mp}^2(Q^2) + \frac{\epsilon}{\tau} G_{Ep}^2(Q^2) , \quad \tau = \frac{Q^2}{4M^2}, \;\; \epsilon^{-1} = 1 + 2(1 + \tau) \tan^2(\theta/2)\,,
\end{equation}
$\theta$ being the scattering angle, $Q^2$ the four momentum transfer, $M$ the proton mass.
For a given $Q^2$ value several measurements at different scattering angles are performed and then the magnetic $G_{Mp}(Q^2)$ and electric $G_{Ep}(Q^2)$ form factors can be simultaneously extracted (see,
e.g. Ref.~\cite{Perdrisat:2006hj}).

Since the beginning of the 90's  new longitudinally polarized
electron beams became available. From that time, experiments on the
elastic scattering of polarized electrons on unpolarized or
polarized target started. In the experiments on the scattering of
polarized electron on unpolarized proton target ($\vec{e}p\to
e\vec{p}$) the transverse and longitudinal polarization of the
recoil protons were measured. The ratio of these quantities is
proportional (in Born approximation) to the form factors ratio
$\mu_p G_{Ep}/G_{Mp}$. In the experiments on the scattering of
polarized electrons on a polarized target ($\vec{e}\vec{p}\to ep$) the
asymmetry was measured. This observable is also a function of the
ratio of the electric and magnetic form factors.

It turned out that the ratio  $\mu_p G_{Ep}/G_{Mp}$ obtained from
polarization
measurements~\cite{Jones:1999rz,Gayou:2001qd,Punjabi:2005wq}, for
values of the four momentum transfer  above 3~GeV$^2$, is in
disagreement with the same quantity evaluated with the proton form
factors extracted with the Rosenbluth separation technique.

An obvious remark is that the Rosenbluth separation method works
less efficiently at large $Q^2$, the reason being that with
increasing $Q^2$ the term $\epsilon G_{Ep}^2/\tau$ becomes
significantly smaller than $G_{Mp}^2$ [see
Eq.(\ref{sigma_reduced})]. Moreover it turns out that even a tiny
additional correcting term to the reduced cross section (of the
order of a few percent of the total $\sigma_R$) can significantly
affect the extracted value of $G_{Ep}$ and hence the form factor
ratio $\mu_p G_{Ep}/G_{Mp}$. For most cross section data the
``classical'' radiative corrections have been applied
\cite{Mo:1968cg}. This however could not solve the problem of the
inconsistency between the form factor ratio obtained from Rosenbluth
technique and polarization measurements.

The possibility to solve this problem by taking into account the
contribution of the two-photon exchange corrections (TPE) was
considered in different papers (see e.g. \cite{Guichon:2003qm}).
These corrections are natural candidate for solving the above
outlined inconsistency. Indeed, at large $Q^{2}$ the contribution of
TPE corrections to the reduced cross section is of the order of
$\epsilon G^{2}_{Ep}/\tau$.

It has been shown that, by including TPE corrections in the data
analysis, one can extract from the  Rosenbluth separation proton
electromagnetic form factors which are in agreement with the results
of polarization
measurements~\cite{Guichon:2003qm,Blunden:2003sp,Chen:2004tw,Afanasev:2005mp}.
It has been also shown that TPE corrections to the ratio of the
transverse and longitudinal polarization of the recoil protons are
smaller than experimental
errors~\cite{Guichon:2003qm,Borisyuk:2007re} and can be, at present,
neglected.
\begin{figure}
\centering{\includegraphics[width=0.65\textwidth]{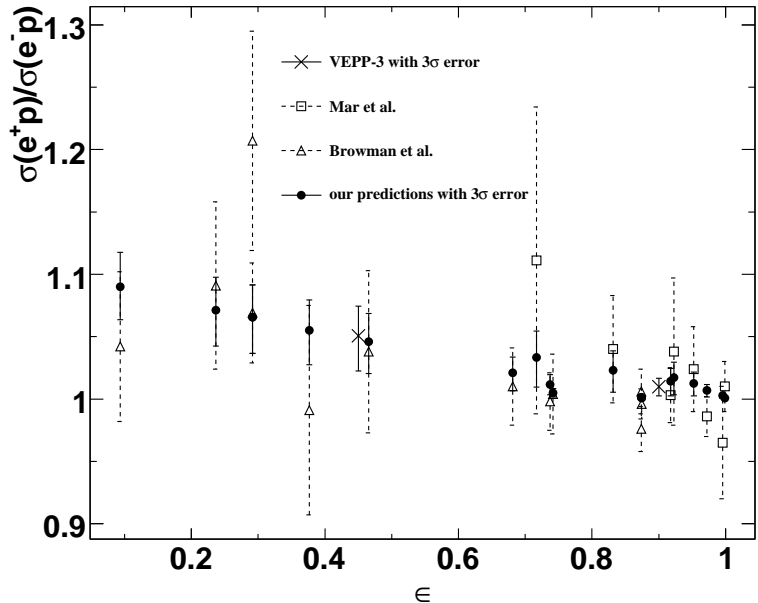}
\caption{The ratio, Eq.~(\ref{ratio_ep}). The open squares and triangles
denote data points from Refs. \cite{Mar:1968qd} and \cite{Browman65}
(the experimental error bars are denoted by dotted intervals). By
black circles our predictions for \cite{Mar:1968qd} and
\cite{Browman65} data are presented together with $3\sigma$ error
bars (denoted by solid intervals). The crosses present our prediction for two points which
are expected to be measured at the VEPP-3 experiment
(together with $3\sigma$ error bars). \label{fig_ratio_ep}} }
\end{figure}

The above problem has been extensively studied by many authors from
the phenomenological and theoretical point of view (for a recent
review see, e.g. Ref.~\cite{Carlson:2007sp}). On the theoretical
side the major difficulty is to take into account properly
intermediate hadronic states, which contribute to the TPE
amplitude~\cite{Kondratyuk:2005kk}. However, the two-photon exchange
contribution can be parameterized phenomenologically and the
corresponding parameters can be determined from the experimental
data~\cite{Borisyuk:2007re,Arrington:2003ck,Arrington:2004ae,Chen:2007ac}.

In the elastic electron-proton scattering process the leading TPE contribution is given by the
interference between the Born and the TPE amplitudes. As a consequence the correcting term is
proportional to the real (dispersive) part of the TPE amplitude and turns out to be odd
in the charge of the projectile. Hence the TPE correction has opposite signs for electron-proton and
positron-proton scattering. For this reason the measurement of the ratio
\begin{equation}
\label{ratio_ep} R_{e^+/e^-}(Q^2,\epsilon) \equiv \frac{\sigma(e^+p
\to e^+p)}{\sigma(e^-p \to e^-p)}
\end{equation}
provides a unique and model-independent possibility to determine the magnitude of the TPE correction.

\begin{figure}
\centering{\includegraphics[width=0.65\textwidth]{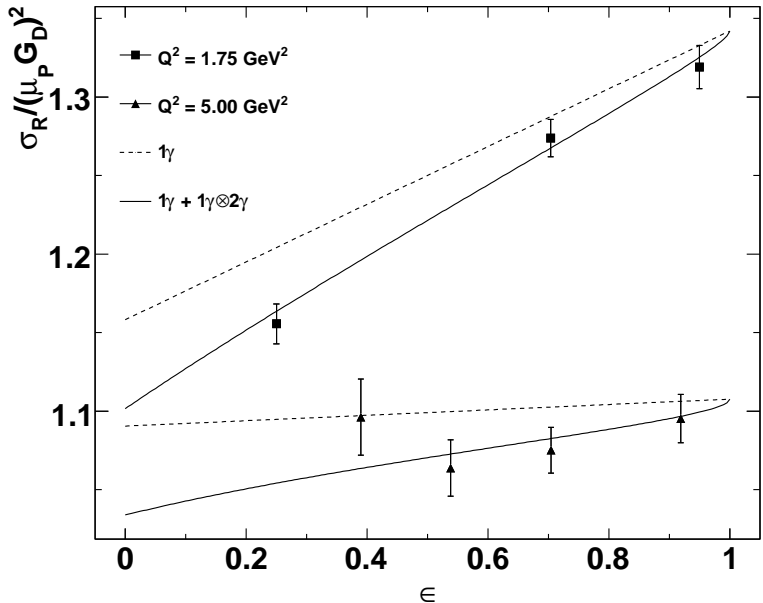}
\caption{Plots of  $\sigma_R/(\mu_p G_D)^2$ for $Q^2=1.75$ and
$5.00$ ~GeV$^2$. The data points are taken from
Ref.~\cite{Andivahis:1994rq}. The solid line denotes the
reduced cross sections computed with the TPE dispersive term, given
by Eq.~(\ref{tpe_term}). The dashed lines correspond to
the same cross sections in Born approximation (i.e. without the TPE
correction). \label{fig_sigmar}} }
\end{figure}

Another possibility to measure the TPE effect is by observing the non-linearity in $\epsilon$
of the Rosenbluth data (see e. g. Fig. \ref{fig_sigmar}),
which is implicated by the symmetry properties of the interaction \cite{Rekalo:2003xa}.
However, up to now such nonlinearities are not visible \cite{Arrington:2003ck,TomasiGustafsson:2004ms,Tvaskis:2005ex}.
Accurate measurements of new Rosenbluth data in a wide
$\epsilon$ range are required to study this effect.

Measurements of the ratio (\ref{ratio_ep}) were performed in the
60's. In Fig.~\ref{fig_ratio_ep} some of the SLAC
data~\cite{Mar:1968qd,Browman65} for the ratio $ R_{e^+/e^-} $ are
shown.
One can see that the experimental uncertainties
do not allow to draw any definite conclusion on the presence of TPE corrections,
since the data are compatible with $ R_{e^+/e^-} = 1 $.

Two new experiments are under preparation, which will measure the elastic electron(positron)-proton scattering
cross sections with higher accuracy. One of them will take place at the VEPP-3 storage ring~\cite{Arrington:2004hk}, the approximate beam energy being around 1.6~GeV. Measurements at $\epsilon\approx 0.90$ and $0.45$ with $Q^2\approx 0.3$ and $1.5$~GeV$^2$, respectively, are proposed. The second experiment was proposed at
JLab~\cite{JLab_experiment}. Here a wide range of $\epsilon \in(0.1,0.9)$ will be explored, with
$Q^2$ in the range $(0.5, 3)$~GeV$^2$.

In this letter we give the expected value for the ratio $ R_{e^+/e^-} $, in the kinematical range accessible to the above mentioned two new experiments. Our results are determined from  the global fit of the $ep$ elastic data~\cite{Alberico:2008sz}. We provide our predictions with the corresponding uncertainties, which are also obtained from our global fit.

\section{Dispersive two-photon exchange contribution}

In Ref.~\cite{Alberico:2008sz} we performed a global fit of the $ep$ elastic cross section and polarization data. The cross section data were fitted by taking into account a TPE correcting term, which was added to the reduced cross section, according to Ref.~\cite{Chen:2007ac}, as follows:
\begin{equation}
\sigma_R(Q^2, \epsilon) \to \sigma_{R,1\gamma}(Q^2, \epsilon) +
\delta_{1\gamma\otimes2\gamma}(Q^2,\epsilon),
\end{equation}
where $\delta_{1\gamma\otimes2\gamma}$ is the TPE correction, due to one- and two-photon interference.
On the other hand,
as mentioned in the Introduction,
the polarization data are practically not affected by TPE corrections.
A decrease of the cross section due to a negative TPE correction allowed us to obtain form factors whose ratio is in agreement
with the  polarization data.
Since the polarization data are necessary for the extraction of the TPE correction through the comparison with the
cross section data,
the global fit in Ref.~\cite{Alberico:2008sz} was limited to the kinematical range of the polarization data.

The dependence on $\epsilon$ of $\delta_{1\gamma\otimes2\gamma}(Q^2,\epsilon)$ is constrained by the requirements
of charge conjugation and crossing symmetry~\cite{Rekalo:2003xa}. As a consequence,
$\delta_{1\gamma\otimes2\gamma}$ should satisfy the relation
\begin{equation}
\delta_{1\gamma\otimes2\gamma}(Q^2,-y) = - \delta_{1\gamma\otimes2\gamma}(Q^2,y),
\end{equation}
where
\begin{equation}
y=\sqrt{\frac{1-\epsilon}{1+\epsilon}}.
\end{equation}
\begin{figure}
\centering{\includegraphics[width=0.95\textwidth]{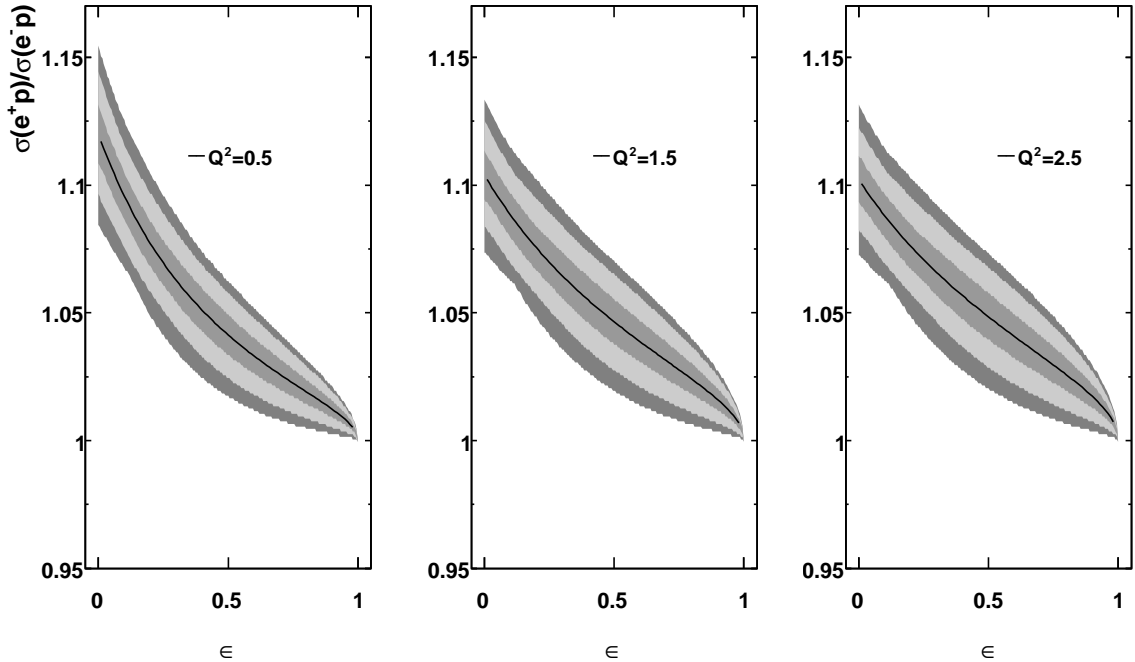}
\caption{The ratio  $ R_{e^+/e^-} $  computed
for three values of $Q^2$ (0.5, 1.5 and 2.5~GeV$^2$)
which will be explored in the JLab experiment E04-116.
The shadowed regions denote $1\sigma$ (the most inner region), $2\sigma$ and
$3\sigma$ (the most external region) calculated uncertainties.
\label{fig_ratio_ep2}}
}
\end{figure}

Based on the above properties Chen~\textit{at al.}~\cite{Chen:2007ac} proposed two parameterizations of the
TPE term. In both of them the $Q^2$ dependence was given by the dipole form factor
\begin{equation}
G_D(Q^2) = \left(1 + \frac{Q^2}{0.71} \right)^{-2}
,
\end{equation}
which was multiplied by functions of $y$. Such a $Q^2$ dependence is heuristically justified by its simplicity and the effectiveness in the fit of the data
(however, a different $Q^2$ dependence with double poles at $ Q^2 = - 1.5\,\mathrm{GeV}^2 $ and $Q^2 =-0.71\,\mathrm{GeV}^2$
was proposed in Ref.~\cite{TomasiGustafsson:2004ms}).
In Ref.~\cite{Alberico:2008sz} we employed the form
\begin{equation}
\label{tpe_term}
\delta_{1\gamma\otimes2\gamma}(Q^2,y) = G_D^2(Q^2) \left(\alpha y + \beta y^3\right),
\end{equation}
which is one of the functional forms proposed in Ref.~\cite{Chen:2007ac}.
The best fit values obtained in Ref.~\cite{Alberico:2008sz} for the parameters $\alpha$ and $\beta$
are\footnote{More precisely these values refer to fit II, but do not significantly differ from the ones of
fit I of the above quoted reference.}
\begin{equation}
\label{par-glo-2-tpe}
\alpha = -0.36 \pm 0.09, \quad \beta = -0.08 \pm 0.09
\,,
\end{equation}
with 1$\sigma$ uncertainties
given by the square-roots of the diagonal elements of the covariance matrix\footnote{
The covariance matrix
and the table of $\chi^2$ values in the parameter space
are given in
\url{http://www.nu.to.infn.it/pap/08/ff/ff.php}.
}.
In the following the uncertainties are evaluated using the correlated systematic uncertainties
of the parameters obtained in Ref.~\cite{Alberico:2008sz}
through a standard least-squares analysis of the data.

In Fig.~\ref{fig_sigmar} we plot the reduced cross
sections (divided by $\mu_p^2 G_D^2$), computed for two $Q^2$ values, 1.75 and
5~GeV$^2$: the solid lines correspond to our global fit~\cite{Alberico:2008sz}, including the TPE term.
We recall that the goodness of fit (GoF) was GoF = 71\%.
By comparing the lines with and without TPE correction, one can see
that the estimated TPE correction is small (of the order of a few percent), the major
effect appearing at small $\epsilon$ values (which correspond to the largest
$y$ values).

We now discuss our prediction for the ratio $ R_{e^+/e^-} $, which was
already shown in Fig.~\ref{fig_ratio_ep} together with the old SLAC data. In this figure, for each
 point we computed the ratio (\ref{ratio_ep}) using the proton form
factors and TPE correction term from our global data fit. The predicted points
are plotted with the $3\sigma$ uncertainty. Although our prediction for the ratio is well above one at small epsilon, the SLAC data do not have sufficient accuracy to reveal a significant deviation from unity.

In the same figure we display also the two points which are going to be
measured in the VEPP-3 experiment~\cite{Arrington:2004hk}. For the kinematical conditions of this
experiment we obtained
\begin{eqnarray}
 && R_{e^+/e^-}(Q^2 = 0.3\,\mathrm{GeV}^2,\epsilon = 0.90) =
 1.010_{-0.002,-0.005,-0.007}^{+0.003,+0.006,+0.007},
\nonumber\\
 && R_{e^+/e^-}(Q^2 = 1.5\,\,\mathrm{GeV}^2, \epsilon = 0.45) =
1.051_{-0.009,-0.019,-0.028}^{+0.009,+0.019, +0.024}, \nonumber
\end{eqnarray}
where the $1\sigma$, $2\sigma$ and $3\sigma$ errors are indicated.
The predicted ratios are clearly above one, by more than $3\sigma$.

Our predictions for the ratio $ R_{e^+/e^-} $ for the kinematical conditions of the E04-116 JLab
experiment~\cite{JLab_experiment} are presented
in Fig.~\ref{fig_ratio_ep2}. We plot $ R_{e^+/e^-} $
for three $Q^2$ values: 0.5, 1.5 and 2.5~GeV$^2$.
The $1\sigma$, $2\sigma$ and $3\sigma$ uncertainties are denoted by shadowed areas.
In all the considered cases the predictions are above one even by more than $3\sigma$.

In both experiments the foreseen accuracy of measurements will be sufficient to reveal the predicted
deviation from unity of $R_{e^{+}/e^{-}}$. For the VEPP-3 experiment the systematic uncertainty is
estimated to be below 0.3\%, while for the JLab project it is predicted to be smaller than 1.0\%. The
statistical errors for the first experiment are expected to be around 1\%,  while for the second project
the statistical uncertainties are foreseen to be smaller than $1\div2\%$
(see Fig. 34 of Ref. \cite{JLab_experiment}).

\begin{figure}
\centering{\includegraphics[width=0.55\textwidth]{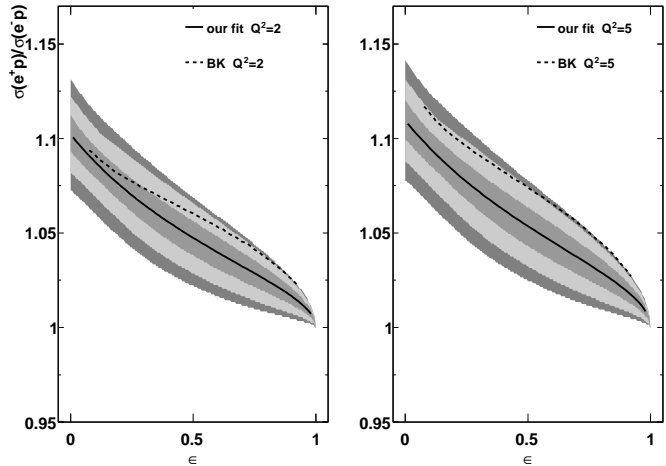}
\caption{The ratio  $ R_{e^+/e^-} $  computed for
$Q^2=2$ and $5$~GeV$^2$, compared with the corresponding predictions in Ref.~\cite{Borisyuk:2008db}.
The shadowed regions denote $1\sigma$ (the most inner region), $2\sigma$ and
$3\sigma$ (the most external region) calculated uncertainties.
\label{fig_ratio_ep4}}
}
\end{figure}

In Fig. \ref{fig_ratio_ep4} we compare our predictions
for $R_{e^{+}/e^{-}}$ with those computed with pQCD in Ref.~\cite{Borisyuk:2008db}.
Since our fit of elastic $ep$ data is valid for
$Q^2<6$~GeV$^2$, we consider only the curves in Fig.~4 of Ref.~\cite{Borisyuk:2008db}
corresponding to $Q^2=2$ and $5$~GeV$^2$.
Our predictions for $R_{e^{+}/e^{-}}$ are systematically lower
than those in Ref.~\cite{Borisyuk:2008db},
but there is an agreement at the $2\sigma$ level.

\section{Conclusions}

As it is well known, the ratio of the electric and magnetic form factors of the proton obtained from the
measurement of the polarization of the recoil proton in the scattering of polarized electrons on unpolarized
protons and that obtained from the measurement of the cross section in the elastic $e^{-}p$-scattering are
not compatible with each other. It was suggested in several papers
\cite{Guichon:2003qm,Blunden:2003sp,Chen:2004tw,Afanasev:2005mp} that this discrepancy can be resolved if one
includes two-photon-exchange corrections.

If the hypothesis of the importance of the two-photon exchange term is correct, then, due to the interference
of the one-photon and two-photon amplitudes, the cross sections in $e^{-}p$ and $e^{+}p$ elastic scattering
are different.

These cross sections were measured in the past at SLAC \cite{Andivahis:1994rq}, however the large errors in
this experiment did not allow to draw definite conclusions about the TPE term. Two new high-precision
experiments on the measurement of the ratio $R_{e^{+}/e^{-}}$ of the cross sections of $e^{\pm}p$ elastic
scattering are now at preparation at the VEPP-3 storage ring \cite{Arrington:2004hk} and at JLab
\cite{JLab_experiment}.

Having in mind these experiments, in this letter we performed the calculation of the ratio $R_{e^{+}/e^{-}}$
in the ranges of $Q^{2}$ and $\epsilon$ which will be covered by these future measurements. We used the
electromagnetic form factors of the proton and the parameters of the TPE term which were obtained from our
global fit \cite{Alberico:2008sz} of the data on the measurement of the cross section and polarization
effects in elastic  $e^{-}p$ scattering. We have shown that the ratio $R_{e^{+}/e^{-}}$ is significantly
different from unity even if we take into account the $3\sigma$ uncertainty, which we obtained from the fit.

\section*{Acknowledgements}

S.~M.~B. has been supported by funds of the Munich Cluster of Excellence
(Origin and Structure of the Universe), the DFG (Transregio 27: Neutrinos and
Beyond) and by INFN Sezione di Torino.

K.~M.~G. was supported by WWS project founds.


\begin{thebibliography}{39}
\bibitem{Perdrisat:2006hj}
  C.~F.~Perdrisat, V.~Punjabi and M.~Vanderhaeghen,
  %``Nucleon electromagnetic form factors,''
  Prog.\ Part.\ Nucl.\ Phys.\  {\bf 59} (2007) 694.
%  [arXiv:hep-ph/0612014].


%%% Polarization Data %%%%%%
%Jon00
\bibitem{Jones:1999rz}
  M.~K.~Jones {\it et al.}  [Jefferson Lab Hall A Collaboration],
%  \textit{G(E(p))/G(M(p)) ratio by polarization transfer in  e(pol.) p --> e p(pol.)},
  Phys.\ Rev.\ Lett.\  {\bf 84} (2000) 1398.
%  [arXiv:nucl-ex/9910005].

%Gay02
\bibitem{Gayou:2001qd}
  O.~Gayou {\it et al.}  [Jefferson Lab Hall A Collaboration],
%  \textit{Measurement of G(E(p))/G(M(p)) in e(pol.) p --> e p(pol.) to Q2 =
%  5.6-GeV2},
  Phys.\ Rev.\ Lett.\  {\bf 88} (2002) 092301.
%  [arXiv:nucl-ex/0111010].


%Pun05
\bibitem{Punjabi:2005wq}
  V.~Punjabi {\it et al.},
  %\textit{Proton elastic form factor ratios to Q**2 = 3.5-GeV**2 by polarization transfer},
  Phys.\ Rev.\  C {\bf 71} (2005) 055202
  [Erratum-ibid.\  C {\bf 71} (2005) 069902].
%  [arXiv:nucl-ex/0501018].

\bibitem{Mo:1968cg}
  L.~W.~Mo and Y.~S.~Tsai,
  %``Radiative Corrections To Elastic And Inelastic E P And Mu P Scattering,''
  Rev.\ Mod.\ Phys.\  {\bf 41} (1969) 205.

%TPE at elastic scattering
\bibitem{Guichon:2003qm}
  P.~A.~M.~Guichon and M.~Vanderhaeghen,
  %``How to reconcile the Rosenbluth and the polarization transfer method in
  %the measurement of the proton form factors,''
  Phys.\ Rev.\ Lett.\  {\bf 91} (2003) 142303.
%  [arXiv:hep-ph/0306007].

\bibitem{Blunden:2003sp}
  P.~G.~Blunden, W.~Melnitchouk and J.~A.~Tjon,
  %``Two-photon exchange and elastic electron proton scattering,''
  Phys.\ Rev.\ Lett.\  {\bf 91} (2003) 142304.
%  [arXiv:nucl-th/0306076].

\bibitem{Chen:2004tw}
  Y.~C.~Chen, A.~Afanasev, S.~J.~Brodsky, C.~E.~Carlson and M.~Vanderhaeghen,
  %``Partonic calculation of the two-photon exchange contribution to elastic
  %electron proton scattering at large momentum transfer,''
  Phys.\ Rev.\ Lett.\  {\bf 93} (2004) 122301.
%  [arXiv:hep-ph/0403058].



\bibitem{Afanasev:2005mp}
  A.~V.~Afanasev, S.~J.~Brodsky, C.~E.~Carlson, Y.~C.~Chen and M.~Vanderhaeghen,
  %``The two-photon exchange contribution to elastic electron nucleon
  %scattering at large momentum transfer,''
  Phys.\ Rev.\  D {\bf 72} (2005) 013008.
%  [arXiv:hep-ph/0502013].



\bibitem{Borisyuk:2007re}
  D.~Borisyuk and A.~Kobushkin,
  %``Phenomenological analysis of two-photon exchange effects in proton form
  %factor measurements,''
  Phys.\ Rev.\  C {\bf 76} (2007) 022201.
%  [arXiv:hep-ph/0703220].


\bibitem{Carlson:2007sp}
  C.~E.~Carlson and M.~Vanderhaeghen,
  %``Two-photon physics in hadronic processes,''
  Ann.\ Rev.\ Nucl.\ Part.\ Sci.\  {\bf 57} (2007) 171.
%  [arXiv:hep-ph/0701272].


\bibitem{Kondratyuk:2005kk}
  S.~Kondratyuk, P.~G.~Blunden, W.~Melnitchouk and J.~A.~Tjon,
  %``Delta resonance contribution to two-photon exchange in electron proton
  %scattering,''
  Phys.\ Rev.\ Lett.\  {\bf 95} (2005) 172503.
%  [arXiv:nucl-th/0506026].


\bibitem{Arrington:2003ck}
  J.~Arrington,
  %``Evidence for two-photon exchange contributions in electron proton and
  %positron proton elastic scattering,''
  Phys.\ Rev.\  C {\bf 69} (2004) 032201.
%  [arXiv:nucl-ex/0311019].

\bibitem{Arrington:2004ae}
  J.~Arrington,
  %``Extraction of two-photon contributions to the proton form factors,''
  Phys.\ Rev.\  C {\bf 71} (2005) 015202.
%  [arXiv:hep-ph/0408261].

\bibitem{Chen:2007ac}
  Y.~C.~Chen, C.~W.~Kao and S.~N.~Yang,
  %``Is there model-independent evidence of the two-photon-exchange effect in
  %the electron-proton elastic scattering cross section?,''
  Phys.\ Lett.\  B {\bf 652} (2007) 269.
%  [arXiv:nucl-th/0703017].

\bibitem{Rekalo:2003xa}
  M.~P.~Rekalo and E.~Tomasi-Gustafsson,
  %``Model independent properties of two-photon exchange in elastic electron
  %proton scattering,''
  Eur.\ Phys.\ J.\  A {\bf 22} (2004) 331.
%  [arXiv:nucl-th/0307066].


%\bibitem{Qattan:2004ht}
%  I.~A.~Qattan {\it et al.},
%  %\textit{Precision Rosenbluth measurement of the proton elastic form factors},
%  Phys.\ Rev.\ Lett.\  {\bf 94} (2005) 142301.

\bibitem{TomasiGustafsson:2004ms}
  E.~Tomasi-Gustafsson and G.~I.~Gakh,
  %``Search for evidence of two photon contribution in elastic electron  proton
  %data,''
  Phys.\ Rev.\  C {\bf 72} (2005) 015209.
  %%CITATION = PHRVA,C72,015209;%%

\bibitem{Tvaskis:2005ex}
  V.~Tvaskis, J.~Arrington, M.~E.~Christy, R.~Ent, C.~E.~Keppel, Y.~Liang and G.~Vittorini,
  %``Experimental constraints on non-linearities induced by two-photon  effects
  %in elastic and inelastic Rosenbluth separations,''
  Phys.\ Rev.\  C {\bf 73} (2006) 025206.
%  [arXiv:nucl-ex/0511021].


%%%%%%  data e^+p/e^-p
\bibitem{Mar:1968qd}
  J.~Mar {\it et al.},
  %``A Comparison Of Electron - Proton And Positron - Proton Elastic Scattering
  %At Four Momentum Transfers Up To 5.0-Gev/C**2,''
  Phys.\ Rev.\ Lett.\  {\bf 21} (1968) 482.


\bibitem{Browman65} A. Browman, F. Liu, and C. Schaerf, Phys.\ Rev.\ 139 B1079 (1965).


\bibitem{Arrington:2004hk}
  J.~Arrington {\it et al.},
  \textit{Two-photon exchange and elastic scattering of electrons / positrons on  the
  proton. (Proposal for an experiment at VEPP-3)},
  arXiv:nucl-ex/0408020.

\bibitem{JLab_experiment} Jefferson Lab experiment E04-116,
\textit{Beyond the Born Approximation: A Precise Comparison of
$e^+p$ and $e^-p$ Scattering in CLAS}, W. K. Brooks, \textit{et
al.}, spokespersons.



\bibitem{Alberico:2008sz}
  W.~M.~Alberico, S.~M.~Bilenky, C.~Giunti and K.~M.~Graczyk, Phys.\ Rev.\  C {\bf 79} (2009) 065204.




\bibitem{Andivahis:1994rq}
  L.~Andivahis {\it et al.},
  %``Measurements of the electric and magnetic form-factors of the proton from Q**2 = 1.75-GeV/c**2 to 8.83-GeV/c**2,''
  Phys.\ Rev.\  D {\bf 50} (1994) 5491.

\bibitem{Borisyuk:2008db}
  D.~Borisyuk and A.~Kobushkin,
  %``Perturbative QCD predictions for two-photon exchange,''
  Phys.\ Rev.\  D {\bf 79}, (2009) 034001.


%\bibitem{Blunden:2005ew}
%  P.~G.~Blunden, W.~Melnitchouk and J.~A.~Tjon,
%  %``Two-photon exchange in elastic electron nucleon scattering,''
%  Phys.\ Rev.\  C {\bf 72}, 034612 (2005).

%\bibitem{Borisyuk:2006uq}
%  D.~Borisyuk and A.~Kobushkin,
%  %``Two-photon exchange at low Q**2,''
%  Phys.\ Rev.\  C {\bf 75}, 038202 (2007).

\end{thebibliography}
\end{document}